\documentclass[aps,prb,floatfix,superscriptaddress,noshowpacs,twocolumn]{revtex4-2}
\usepackage{graphicx,amsmath,amssymb,epstopdf,wasysym}
\usepackage[space]{grffile}
\usepackage{hyperref}
\usepackage[dvipsnames]{xcolor}
\usepackage{braket,amsfonts}
\usepackage{dsfont}
\usepackage{float} 

\allowdisplaybreaks

\newcommand{\be}{\begin{equation}}
\newcommand{\ee}{\end{equation}}
\newcommand{\bga}{\begin{gather}}
\newcommand{\ega}{\end{gather}}
\newcommand{\bea}{\begin{eqnarray}}
\newcommand{\eea}{\end{eqnarray}}
\newcommand{\dagga}{{\phantom{\dagger}}}
\newcommand{\bR}{\mathbf{R}}

\newcommand{\bM}{\mathbf{M}}

\newcommand{\bk}{\mathbf{k}}

\newcommand{\dis}{\displaystyle}

\newcommand{\up}{\uparrow}
\newcommand{\down}{\downarrow}
\newcommand{\fract}[2]{\frac{\dis \;#1\;}{\dis \;#2\;}}

\newcommand{\eqn}[1]{(\ref{#1})}

\newcommand{\ep}{{\epsilon}}

\newcommand{\bw}{\begin{widetext}}
\newcommand{\ew}{\end{widetext}}
\newenvironment{eqs}%
{\begin{equation} \begin{aligned}}%
{\end{aligned} \end{equation} }
\newcommand{\beal}{\begin{eqs}}
\newcommand{\eal}{\end{eqs}}
\newcommand{\bd}[1]{{\boldsymbol{#1}}}

\newcommand{\bealn}{\beal\nonumber}

\begin{document}
\title{Exciton condensation driven by bound states of Green's functions zeros}
\author{Ivan Pasqua}
\affiliation{International School for Advanced Studies (SISSA), Via Bonomea 265, I-34136 Trieste, Italy} 

\author{Andrea Blason}
\affiliation{International School for Advanced Studies (SISSA), Via Bonomea 265, I-34136 Trieste, Italy}

\author{Michele Fabrizio}
\affiliation{International School for Advanced Studies (SISSA), Via Bonomea 265, I-34136 Trieste, Italy} 

\begin{abstract} 
The interaction driven transition between quantum spin-Hall and Mott insulators
in the Bernevig, Hughes and Zhang model is studied by 
dynamical cluster approximation, and found to be accompanied by the emergence of Green's function zeros already in the quantum spin-Hall regime. The non-trivial interplay between Green's function poles and zeros leads to an exotic quantum spin-Hall insulator exhibiting two chiral branches of edge Green's function poles and one of zeros. When symmetry breaking is allowed, a non-topological excitonic insulator is found to intrude between quantum spin-Hall and Mott insulators. We find evidence that excitons in the Mott insulator, which become soft at the transition to the excitonic insulator, are actually bound states between valence and conduction bands of Green's function zeros, rather than between lower and upper Hubbard bands.
\end{abstract}

\maketitle 

\section*{Introduction}
Upon increasing the electron-electron interaction strength, models of topological band insulators (TBIs) can be driven into supposedly-trivial, non-symmetry breaking correlated insulators 
(CIs) without the expected closing at the transition of the single-particle gap. Several ways have been found that allow circumventing the occurrence of a metal point between TBI and CI. The simplest and most direct is a discontinuous character of the 
transition \cite{Amaricci2015PRL,Amaricci2016PRB}, but there seem to exist 
other more subtle routes connecting a TBI to a CI without crossing a metal.\\ 
For instance, a new insulating phase can intrude between TBI and CI that spontaneously breaks the symmetry which protects the topology. 
In the Bernevig, Hughes and Zhang (BHZ) model of quantum spin-Hall insulators (QSHIs) \cite{BHZ}, this phase describes a magnetoelectric excitonic insulator, which thus breaks parity and time-reversal \cite{Blason2020PRB,Adriano-PRB2023}. As a matter of fact, exciton condensation is a logical outcome when the single-particle gap gets small around the TBI-CI transition. Indeed, there is evidence of indirect-exciton condensation in monolayer WTe$_2$ \cite{Yazdani-NatPhys2022}, which leads to 
a topological insulator that breaks time-reversal symmetry. \\
A further and rather exotic scenario is that the TBI-to-CI transition is direct and continuous, the single-particle gap remains finite and smooth, and, yet, a gapless point emerges within the particle-hole and/or particle-particle spectra. Slagle, You and Xu \cite{Slage-PRB2015} have studied 
an AA-stacked honeycomb-lattice bilayer where each layer is described by a 
Kane-Mele model \cite{Kane&Mele-PRL2005} of a QSHI plus an on-site Hubbard $U$, 
and the two layers are coupled to each other by an antiferromagnetic exchange $J$. 
Upon increasing $U$ at large $J/U=2$, they find a continuous transition from the 
QSHI to a trivial CI that is simply a collection of inter-layer singlets. 
At the transition, the single-particle gap remains finite whereas the spin gap 
as well as the gap to adding/removing two electrons 
vanish. 
This result raises an interesting issue. Indeed, the QSHI phase 
is characterised by a topological invariant $C_\up=-C_\down=\pm 1$, with $C_\sigma$ the Chern number of 
the $\sigma$-band, which is also the winding number 
$W(G_\up)=-W(G_\down)$ of the map 
$(\ep,\bk)\to G_\up(i\ep,\bk) \in GL(2,\mathbb{C})$ \cite{Zhang-PRX2012}, where $G_\up(i\ep,\bk)$ is the spin-up Green's function in Matsubara frequencies $\ep$, and $\bk$ the two-dimensional momentum. Remarkably, the winding number retains the same quantised value on both sides of 
the transition \cite{Slage-PRB2015}, which, in turn, implies that the topological bands of poles of the QSHI Green's function determinant at real, negative  frequencies transform into 
topological bands of zeros in the CI \cite{Gurarie-zeros-PRB2011,Essing&Gurarie-PRB2011,Slage-PRB2015,Andrea-Zeri-PRB2023,Giorgio-NatComm2023}.
We emphasise that the topological invariant  $W(G_\up)-W(G_\down)$ corresponds to the quantised spin-Hall conductance $C_\up-C_\down$ in the QSHI, but it does not in the CI \cite{Slage-PRB2015,He-PRB2016-II,Andrea-Zeri-PRB2023}. \\
How the transmutation of topological poles into topological zeros occurs across the transition is not known but in rather artificial cases \cite{Phillips-PRL2023}. It is also unclear whether such change of analyticity bears    
any signature of the vanishing spin and charge-$2e$ gaps, even though it is tempting to argue that their closing despite the finite charge-$e$ gap corresponds to a Dirac-like touching of the bands of zeros, thus the emergence of a Luttinger 
surface \cite{mio-Mott}, made just of two points in the specific case.  \\
\noindent
In this work we try to shed some light on those issues studying the transition in the BHZ model \cite{BHZ} between QSHI and CI, where the presence of the intermediate excitonic insulator (EI) phase raises further intriguing questions. Indeed, within dynamical mean-field theory (DMFT) \cite{DMFT} the transition upon increasing the interaction strength $U$ between the EI and the Mott correlated insulator (MI) is continuous \cite{Adriano-PRB2023}, which implies the existence of an excitonic mode in the Mott phase that gets soft approaching the transition and condenses beyond it, in the EI. In spite of that, the single-particle gap is sizeable and smooth across the transition \cite{Adriano-PRB2023}. In other words, 
the exciton softening seems to a large extent unrelated to the Mott gap between lower and upper Hubbard bands. Since the latter entail in-gap bands of zeros, it is tempting to argue that the excitons are  
bound states of valence and conduction bands of zeros, a suggestive conjecture that we shall investigate. 

\section{Model and methods} 
\label{Model and methods}
We consider the BHZ model \cite{BHZ} for the QSHI  
phase of HgTe quantum wells. This model involves two spinful Wannier orbitals per unit cell, one that transforms like $s$-orbitals, $\ket{\ell=0,\sigma}\equiv \ket{s \,\sigma}$, $\sigma=\up,\down$ being the spin projection along $z$, and the other like the $J=3/2$, $J_z=\pm 3/2$ spin-orbit 
coupled combinations of $p$-orbitals, 
$\ket{\ell=1,\ell_z=+1,\up}\equiv \ket{p \up}$ and 
$\ket{\ell=1,\ell_z=-1,\down}\equiv \ket{p \down}$.
We introduce two sets of Pauli matrices, $\sigma_a$ and $\tau_a$, $a=0,\dots,3$, with $a=0$ denoting the identity, which act, respectively, in the spin, $\up$ and $\down$, and orbital, $s$ and $p$, spaces.\\
With those definitions, the BHZ tight-binding Hamiltonian on a square lattice 
and in momentum space reads
\be
H_0 = \sum_\bk\,  \Psi^\dagger_\bk  \; \hat H_0(\bk) \;\Psi^\dagga_\bk 
 , \label{BHZ-Ham}
\ee
at density corresponding to two electrons per site, where 
$\Psi^\dagger_\bk = \big(s^\dagger_{\bk\up},
s^\dagger_{\bk\down},
p^\dagger_{\bk\up},
p^\dagger_{\bk\down}\big)$
are four component spinors. 
In this representation, $\hat{H}_0(\bk)$ is the $4\times 4$ matrix
\beal
\hat H_0(\bk) &= \Big(M- t\,\big(\cos k_x+\cos k_y \big)\Big) \,\sigma_0\otimes\tau_3   \\ 
 &\quad +\lambda\,\sin k_x\;\sigma_3\otimes\tau_1 
-\lambda\,\sin k_y\;\sigma_0\otimes\tau_2 \, .\label{H_0}
\eal
Hereafter, we set $t = 1$, the energy unit, $\lambda = 0.3$, half-filled density, i.e., two electrons per site, and, without loss of generality, $M\geq 0$.\\
\noindent
For $M<2$ the Hamiltonian \eqn{H_0} describes a QSHI, otherwise 
a conventional non-topological band insulator. \\
The periodic model \eqn{BHZ-Ham} is invariant under $C_4$, inversion 
$\mathcal{I}$, particle-hole $\mathcal{P}$ and time-reversal $\mathcal{T}$ symmetries, as well as under spin $U(1)$ rotations around the $z$-axis.\\
The on-site Coulomb interaction projected onto the Wannier orbital basis can be written as 
\beal
H_\text{int} &= \sum_\bR\, H_\text{int}(\bR)\,,
\label{interaction}
\eal
where $\bR$ labels the lattice sites and 
\beal
H_\text{int}(\bR) &= 
U_s\,n_{s\bR\up}\,n_{s\bR\down} 
+ U_p\,n_{p\bR\up}\,n_{p\bR\down} \\
&\qquad + V\,n_{s\bR}\,n_{p\bR} + H_\text{dip}(\bR)\,,
\label{interaction-1}
\eal
with $n_{a\bR\sigma}$ the number of spin $\sigma=\up,\down$ electrons 
at site $\bR$ in orbital $a=s,p$, and 
$n_{a\bR}=n_{a\bR\up}+n_{a\bR\down}$.
The term $H_\text{dip}(\bR)$ in \eqn{interaction-1} is the 
the dipole component of the multipole expansion of the Coulomb interaction, 
\beal
&H_\text{dip}(\bR) = \fract{J}{2}\,\bigg\{
\Big(\Psi^\dagger_\bR\,\sigma_0\otimes\tau_1\,\Psi^\dagga_\bR\Big)^2\\
&\qquad \qquad \qquad \qquad+\Big(\Psi^\dagger_\bR\,\sigma_3\otimes\tau_2\,\Psi^\dagga_\bR\Big)^2\bigg\}\\
&\qquad = J\,\Big(s^\dagger_{\bR\up}\,s^\dagger_{\bR\down}\,
p^\dagga_{\bR\down}\,p^\dagga_{\bR\up}
+ p^\dagger_{\bR\up}\,p^\dagger_{\bR\down}\,
s^\dagga_{\bR\down}\,s^\dagga_{\bR\up} \\
&\qquad\qquad\;   + s^\dagger_{\bR\up}\,p^\dagger_{\bR\up}\,s^\dagga_{\bR\up}\,p^\dagga_{\bR\up}
+ s^\dagger_{\bR\down}\,p^\dagger_{\bR\down}\,s^\dagga_{\bR\down}\,p^\dagga_{\bR\down}\Big)\,,
\label{interaction-dipole}
\eal
where $\Psi^\dagga_\bR$ is the Fourier transform in real space of the spinor $\Psi^\dagga_\bk$, 
$\bR$ being a lattice site.
All parameters, $U_s$, $U_p$, $V$ and $J$ 
are positive. The interaction \eqn{interaction-1} enforces Hund's rules when $\text{min}(U_s,U_p)>V$, which we assume hereafter and entails that the lowest energy two-electron configuration  
of $H_\text{int}(\bR)$ is a spin triplet, $S=1$, with $S_z=\pm 1$.\\
\noindent
However, for completeness, we will also consider the case $J<0$. This may occur when the electrons are strongly coupled to an infrared optical mode twofold degenerate by $C_4$, 
\bealn
H_\text{el-ph}(\bR) &= q_{1\bR}\,
\Big(\Psi^\dagger_\bR\,\sigma_0\otimes\tau_1\,\Psi^\dagga_\bR\Big)\\
&\qquad  \qquad+ q_{2\bR}\,\Big(\Psi^\dagger_\bR\,\sigma_3\otimes\tau_2\,\Psi^\dagga_\bR\Big)\,,
\eal
which, integrating out phonons and discarding retardation effects, yields 
an effective interaction of the same form as \eqn{interaction-dipole} but with a
negative coupling constant. When the latter overwhelms the Coulomb exchange, the net effect is a dipole term \eqn{interaction-dipole} with an effective $J<0$.
In this case, the ground state of $H_\text{int}(\bR)$ 
with two electrons is twofold degenerate and comprises the states where the two orbitals are singly occupied and coupled either in a spin-singlet, $S=0$, or in a spin-triplet, $S=1$, with $S_z=0$. \\
To make contact with \cite{Slage-PRB2015}, we will add for $J<0$ the additional term 
\beal
\delta H_\text{dip}(\bR) &= 
-\delta J\,\Big(s^\dagger_{\bR\up}\,p^\dagger_{\bR\down}\,
s^\dagga_{\bR\down}\,p^\dagga_{\bR\up} + H.c.\Big)\,,
\label{interaction-dipole-1}
\eal
with $\delta J>0$, which lowers the energy of the singlet configuration and thus 
stabilises a non-magnetic Mott insulator, actually 
a Van Vleck paramagnet because spin $SU(2)$ is broken. We remark, however, that \eqn{interaction-dipole-1} breaks explicitly the $C_4$ symmetry, $p^\dagga_{\bR\up}\to i\,p^\dagga_{\bR\up}$ and $p^\dagga_{\bR\down}\to -i\,p^\dagga_{\bR\down}$.

\subsection{Leading symmetry breaking channels}
The leading symmetry breaking channels can be identified as the most negative 
bare scattering amplitudes \cite{Blason2020PRB}. To reduce the parameter freedom and thus simplify the analysis, we hereafter take $U_s=U_p=U$. We define the local scattering channels and the 
corresponding bare scattering amplitudes as 
$\Psi^\dagger_\bR\,\sigma_a\otimes \tau_b\,\Psi^\dagga_\bR$ and $\Gamma^0_{ab}$, respectively. \\
For $J>0$, the most negative symmetry breaking amplitudes are \cite{Blason2020PRB}
\begin{itemize}
\item
$\Gamma^0_{30}=-(U+2J)/4$, which tends to stabilise ferro or antiferro magnetic order with local magnetisation parallel or antiparallel to $z$;
\item $\Gamma^0_{11}=\Gamma^0_{21}=-(V+2J)/4$, degenerate by spin $U(1)$, which can drive an exciton condensation with order parameter
\beal
\Delta(\phi) &= \cos\phi\,\langle\,\Psi^\dagger_\bR\,\sigma_1\otimes\tau_1\,
\Psi^\dagga_\bR\,\rangle\\
&\qquad + \sin\phi\,\langle\,\Psi^\dagger_\bR\,\sigma_2\otimes\tau_1\,
\Psi^\dagga_\bR\,\rangle\,,
\label{exciton J>0}
\eal
breaking $\mathcal{I}$, $\mathcal{T}$ and spin $U(1)$ symmetries. 
\end{itemize}
Since $U>V$, the magnetic instability is more likely in the MI. However, as we mentioned, the excitonic instability does emerge between the QSHI and MI. \\
\noindent
For $J<0$, the most negative symmetry breaking amplitudes are instead \cite{Blason2020PRB}
\begin{itemize}
\item
$\Gamma^0_{33}=-(U+2|J|)/4$, which favours an ordering in the Mott phase of the local configurations $s^\dagger_{\bR\up}\,p^\dagger_{\bR\down}$ and $p^\dagger_{\bR\up}\,s^\dagger_{\bR\down}$;
\item $\Gamma^0_{01}=\Gamma^0_{32}=-(V+4|J|)/4$, degenerate by $C_4$, which can drive an exciton condensation. 
\end{itemize}
Adding a weak $\delta J >0$, which we recall breaks $C_4$, stabilises a non-magnetic Mott insulator where each site is locked into the inter-orbital spin-singlet  
\bealn
\fract{1}{\sqrt{2}}\,\Big(s^\dagger_{\bR\up}\,p^\dagger_{\bR\down}+p^\dagger_{\bR\up}\,s^\dagger_{\bR\down}\Big)\ket{0}\,,
\eal
whereas the leading excitonic instability corresponds to the order parameter 
\beal
\Delta &= \langle\,\Psi^\dagger_\bR\,\sigma_0\otimes\tau_1\,
\Psi^\dagga_\bR\,\rangle\,,
\label{exciton J<0}
\eal
that breaks inversion.
\subsection{Methods}
The results that we present hereafter are obtained by the so-called dynamical cluster approximation (DCA) \cite{DCArev}, which consists in partitioning the Brillouin zone into patches and in approximating the self-energy as a piecewise constant function within these patches. The many-body problem is tackled by self-consistently solving coupled quantum impurity models. The impurity solver is implemented within the TRIQS library \cite{triqs} and makes use of a continuous time quantum Monte Carlo algorithm based on a hybridization expansion of the partition function \cite{CTHYB_1, CTHYB_2}. \\
In order to reconstruct a smooth momentum dependence from the piecewise constant self-energy obtained in DCA, one can follow two alternative routes: either periodise the self-energy or its inverse in terms of hopping processes with increasing spatial range, consistently with the space group symmetry and the finite number of patches. The former approximation seems to work better at weak coupling, while the latter, known as \text{cumulant} 
periodisation \cite{Cum_exp}, is more suitable when the self-energy develops singularities in the Brillouin zone, which commonly occurs
near the Mott transition and inside the Mott insulator. \\
In the calculations we employ just two patches centred around $\bd{\Gamma} = (0, 0)$ and $\mathbf{M} = (\pi, \pi)$, which was already computationally expensive since the model has two interacting orbitals per site. We use the same partition of the Brillouin zone as in \cite{Lorenzo-patches-PRB2009}, in which the patch at $\bM$ includes also the $\mathbf{X}=(\pi,0)$ and $\mathbf{Y}=(0,\pi)$ points, but we checked that different partitions do not change appreciably the results. \\
The 2-patch scheme has however a drawback that it is worth highlighting now. The Green's function $\hat{G}(i\ep,\bk)$ in Matsubara frequencies $\ep=(2\ell+1)\,\pi\,T$ with $\ell \in \mathbb{Z} $, is a $2\times 2$ matrix satisfying Dyson's equation 
\beal
\hat{G}(i\ep,\bk)^{-1} &= i\ep - \hat{H}_0(\bk) -\hat{\Sigma}(i\ep,\bk)\,,
\label{Dyson's equation}
\eal
where the self-energy matrix $\hat{\Sigma}(i\ep,\bk)$ encodes all interaction effects. However, $\hat{\Sigma}(i\ep,\bk)$ 
is purely diagonal at $\bk=\bd{\Gamma},\mathbf{M}$ because of inversion symmetry. Therefore, by using just 
the two patches at $\bd{\Gamma}$ and $\mathbf{M}$, we cannot access the 
off-diagonal elements of $\hat{\Sigma}(i\ep,\bk)$ that must exist away from the inversion symmetry points, and which would require a number of patches $\geq 8$ that is numerically intractable for us. In the following, we shall point out features of our results that may be critically affected by this limitation, and argue what could change should we use a larger number of patches. \\
\noindent
Finally, before delving into the analysis of our results, let us recall a recently introduced interpreting scheme \cite{mio-Mott} aimed at providing a quasiparticle description applicable to both weakly and strongly correlated regimes. We write the interacting Green's function matrix on the Matsubara frequency axis as
\beal
\hat{G}(i\epsilon, \mathbf{k}) 
& = \frac{1}{i \epsilon - \hat{H}_{0}(\mathbf{k}) - \hat{\Sigma}(i \epsilon, \mathbf{k}) } \\
& = \hat{A}(\epsilon, \mathbf{k}) \, \frac{1}{i \epsilon - \hat{H}_{*}(\mathbf{k}) \;}\,
 \hat{A}(\epsilon, \mathbf{k})^\dagger \\
& \equiv \hat{A}(\epsilon, \mathbf{k}) \,   G_{*}(i\epsilon, \mathbf{k}) \,\hat{A}(\epsilon, \mathbf{k})^\dagger\, ,
\label{G*}
\eal
where, if we define the hermitian matrices 
\beal
\hat{\Sigma}_1( i\epsilon, \mathbf{k} ) & = \frac{\hat{\Sigma}(i\epsilon, \mathbf{k} ) + \hat{\Sigma}(i\epsilon, \mathbf{k} )^\dagger}{2}\;, \\
\hat{\Sigma}_2( i\epsilon, \mathbf{k} ) & = \frac{\hat{\Sigma}(i\epsilon, \mathbf{k} ) - \hat{\Sigma}(i\epsilon, \mathbf{k} )^\dagger}{2i}\;, 
\eal
since $\hat{\Sigma}(i\ep,\bk)^\dagger=\hat{\Sigma}(-i\ep,\bk)$, 
then
\beal
\hat{Z}( \epsilon, \mathbf{k} ) &= \hat{Z}( -\epsilon, \mathbf{k} )=\bigg( 1 - \frac{ \hat{\Sigma}_2( i\epsilon, \mathbf{k} ) }{\epsilon} \bigg)^{-1}\\
&=\hat{A}(\epsilon, \mathbf{k})^\dagger\,\hat{A}(\epsilon, \mathbf{k})
\;,
\label{Z}
\eal
is a semi positive-definite matrix, while the hermitian \textit{quasiparticle} (QP) Hamiltonian reads
\beal
&\hat{H}_{*}(\epsilon, \mathbf{k} ) = \hat{H}_{*}(-\epsilon, \mathbf{k} )\\
&\quad =\hat{A}(\epsilon, \mathbf{k}) \left( \hat{H}_{0}(\mathbf{k}) + \hat{\Sigma}_1( i\epsilon, \mathbf{k} ) \right) \hat{A}(\epsilon, \mathbf{k})^\dagger\,.
\label{H*}
\eal
In our discussion we will always refer to the low-energy behaviour of the QP Hamiltonian $ \hat{H}_{*}(0, \mathbf{k}) = \hat{H}_{*}(\mathbf{k})$. In the weakly interacting QSHI regime the QP bands practically coincide with the dispersive poles of the Green's function. In the MI instead, the in-gap QP bands are related to the presence of in-gap bands of Green's function zeros on the real axis, i.e., to the roots of 
\bealn
\text{det}\big(G(i\ep\to \varepsilon_\bk,\bk)\big) &=0\,,&
\ep_\bk\in\mathbb{R}\,.
\eal
The QP bands and $-\varepsilon_\bk$ coincide in the limit of very large Mott-Hubbard gap \cite{Andrea-Zeri-PRB2023,Giorgio-NatComm2023}. We will show that such a correspondence is effective even for intermediate coupling, thus facilitating a description of the excitonic instability. \\

\begin{figure}[t!] 
\includegraphics[width=0.5\textwidth]{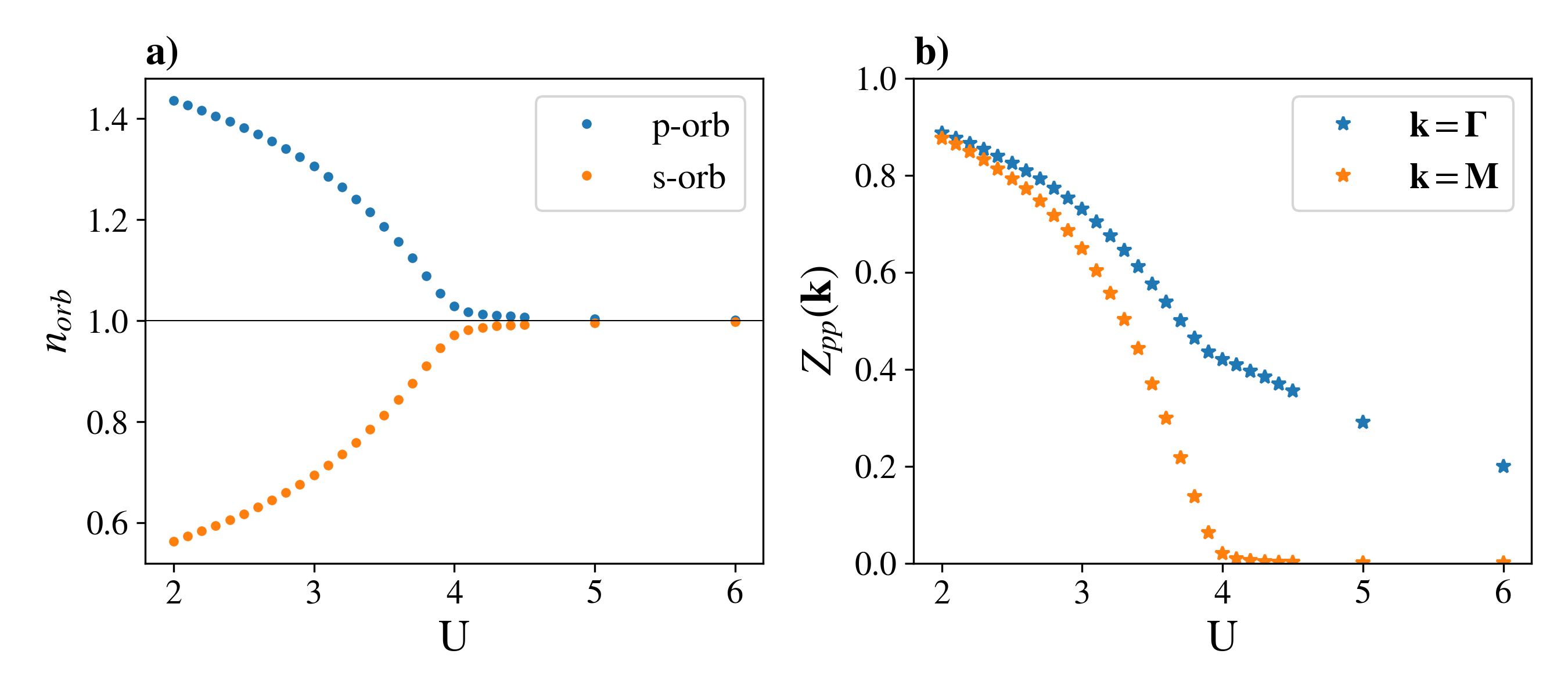}
\caption{\label{fig:Figure1}Panel a): Renormalization of the occupations in the two orbitals across the QSHI-MI transition, at fixed M = 0.9. Deep in the MI phase $n_p - n_s$ decays as $1/U^3$.  Panel b): Quasiparticle residue $\hat{Z}_{pp}(\mathbf{k}) = \hat{Z}_{ss}(\mathbf{k})$ computed from the imaginary part of the self-energy at the first Matsubara frequency. In the $\mathbf{M}$-patch it is strongly suppressed inside the MI, signaling the proximity to a self-energy pole. The change of concavity around $U\sim4$ is identified as the onset of Mottness.}
\end{figure}

\section{The symmetric QSHI-MI transition}
Unless stated otherwise, the numerical results were obtained at $\beta = 50$ and  it was checked their stability upon decreasing the temperature.

\subsection{ The $J > 0$ case}

We set $V = U - 2J$ and $J = U / 4$. The phase diagram in the ($U$, $M$) plane within the DMFT approximation has been extensively discussed with and without excitonic symmetry breaking \cite{Adriano-PRB2023}. First of all, we here address the symmetric QSHI to MI transition, i.e. in absence of spontaneous symmetry breaking (SSB). In single-site DMFT the transition is first order, but we will argue that the inclusion of non-local correlations using the DCA scheme changes the picture. In Fig.~\ref{fig:Figure1} we show the occupation number $n_{orb}$ in the two orbitals as we increase the interaction strength $U$ at fixed $M=0.9$. For weak interaction strength the model is a QSHI adiabatically connected to the non-interacting one and the main contribution of the interaction is a renormalization of the on-site mass difference between the two orbitals and, consequently, of the orbital magnetization $n_p - n_s$:
\beal
M_{\text{eff}}  = M + \frac{1}{4} \text{Tr}[ \sigma_{0} \otimes \tau_{3} \, \hat{\Sigma}(i \epsilon = 0) ]
\eal
where $ \hat{\Sigma}(i \epsilon) = \left( \hat{\Sigma}(i\epsilon, \mathbf{\Gamma}) + \hat{\Sigma}(i \epsilon, \mathbf{M}) \right) / 2$ is the local self-energy.
\begin{figure}[t!] 
\includegraphics[width=0.5\textwidth]{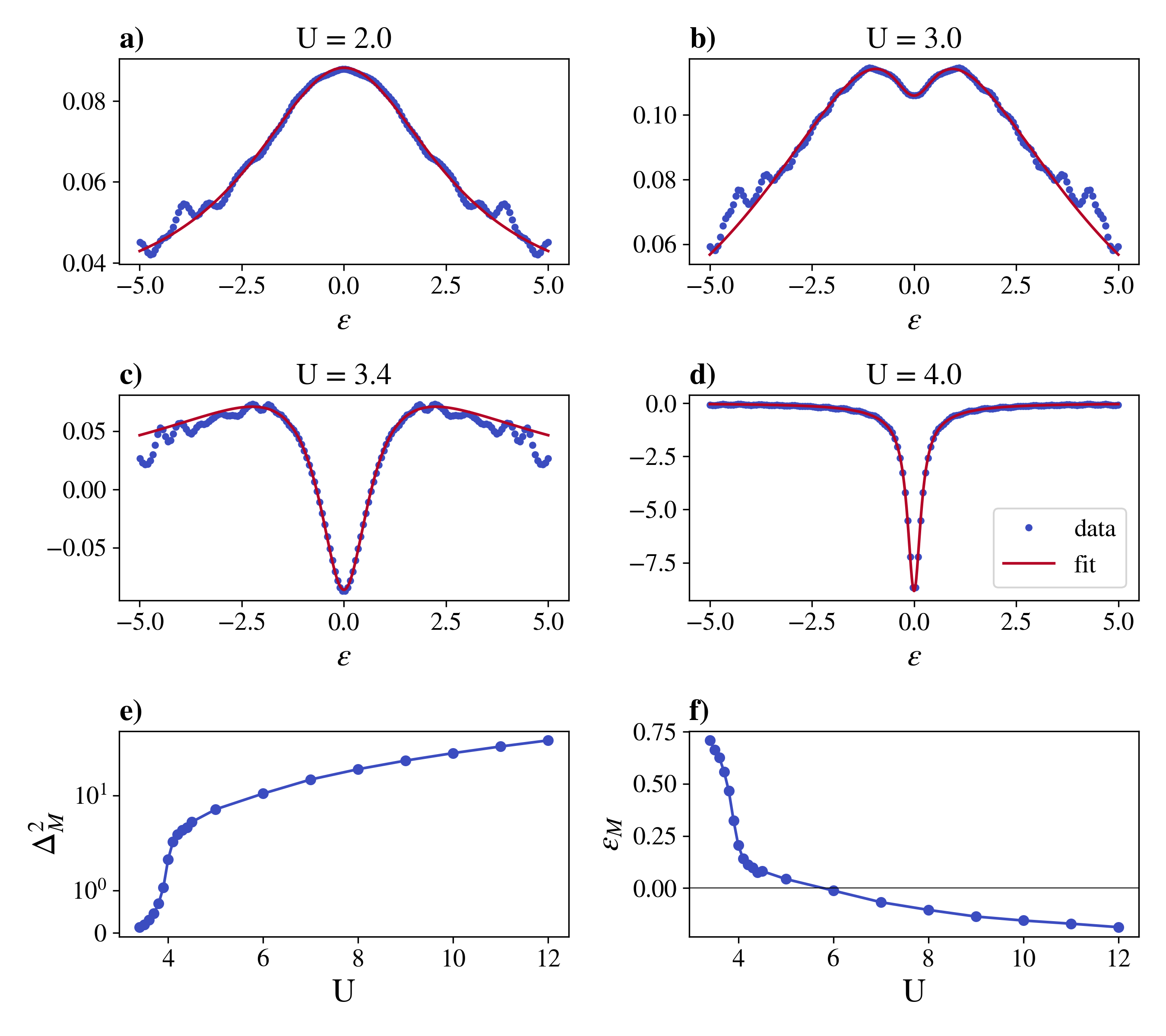}
\caption{\label{fig:Figure2}Panels a, b, c, d): $\Delta \hat{\Sigma}_{pp} ( i \epsilon, \mathbf{M}) = -\hat{\Delta} \Sigma_{ss} ( i \epsilon, \mathbf{M})$ for 4 different values of $U$, at fixed $M=0.9$. For $U \ge 3$ a narrower peak appears and becomes dominant as we increase the interaction strength. There is a good agreement between the data and the fit \eqn{fit}, from which we can extract information about the narrower Lorentzian. Panels e, f): fit parameters of the narrower Lorentzian related with a pole-like structure emerging in the self-energy. $\epsilon_{\mathbf{M}}$ changes sign between $U = 5$ and $U = 6$ through a Luttinger Dirac-like surface in $\mathbf{M}$. }
\end{figure}
\begin{figure}[t!] 
\includegraphics[width=0.5\textwidth]{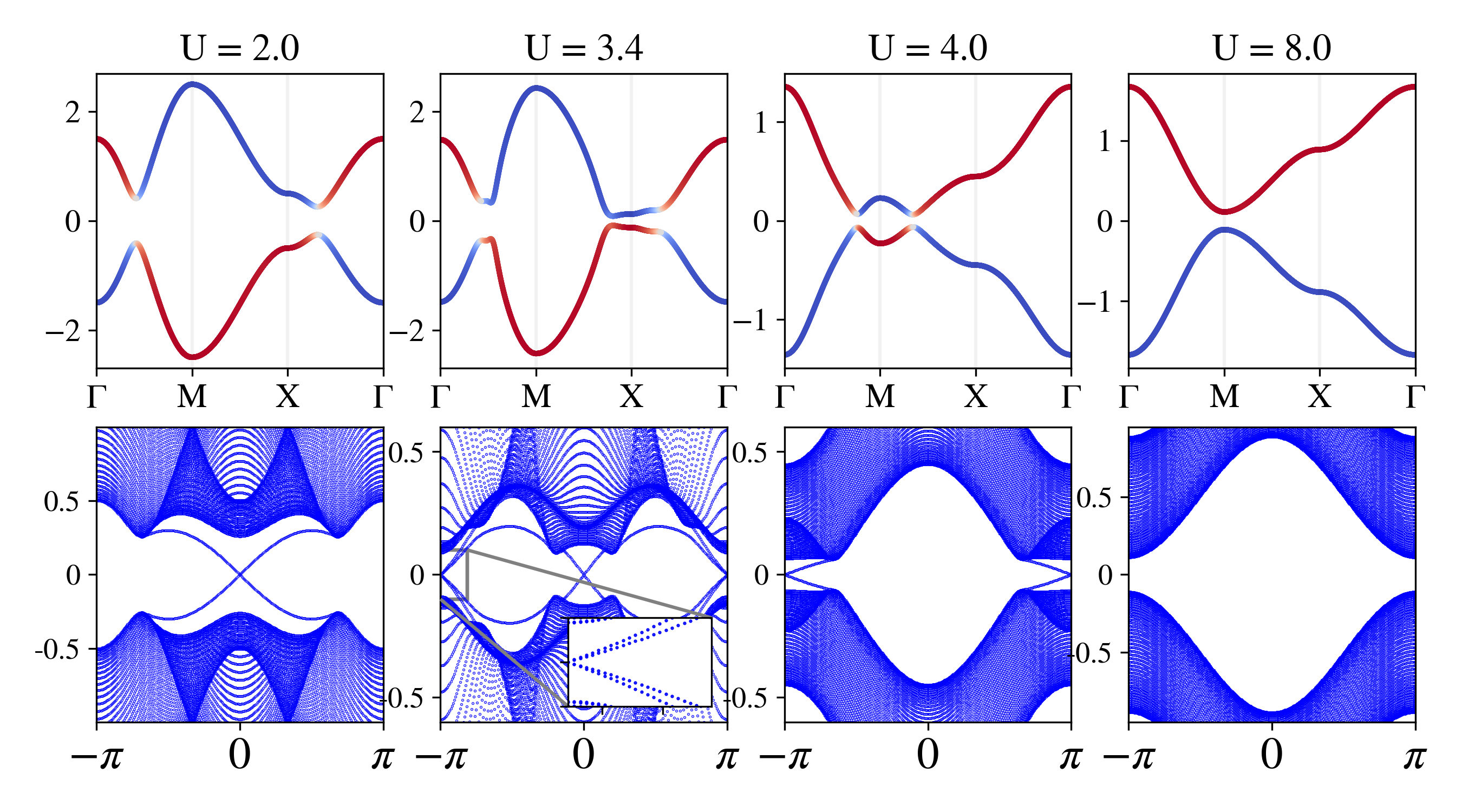}
\caption{\label{fig:Figure3}Using the parameters fitted from the self-energy (see Fig.~\ref{fig:Figure2}) in the toy model \eqn{toy-R} we computed the bands of $\hat{H}_{*}(0, \mathbf{k})$ for 4 different values of $U$, both in PBC (upper panels) and OBC (lower panels). The colours of the lines represent the orbital character of the eigenvectors, blue (red) for the s-(p-)character. For $U=2.0$ the system is adiabatically connected to the non-interacting QSHI. For $U = 3.4$ the topology is unchanged, but in OBC we recognize three pairs of chiral edge states: two associated with Green’s function poles and one with zeros. In the inset we zoomed a region around $\mathbf{k} = - \pi$. For $U=4$  the model turns into a MI that has topological in-gap bands of zeros with winding numbers opposite to the QSHI. Finally for $U=8$ the in-gap zeros are topologically trivial and smoothly connected to the atomic limit. }
\end{figure}
In the strong coupling regime the system is a MI with almost identical occupation in the two orbitals. Indeed, deep in the Mott insulating phase, one expects in perturbation theory that $n_p - n_s$ decays as $1/U^3$. Since $J>0$ enforces Hund's rules, the two electrons localised on each site form a spin triplet state with $S_{z}=\pm1$, thus the acronym hs-MI where hs stands for high-spin. For completeness in Fig.~\ref{fig:Figure1} we show also the quasiparticle residue \eqn{Z} in the two patches computed from the imaginary part of the self-energy at the first Matsubara frequency $i \epsilon_1 = i \pi T $. Differently from the DMFT case, we observe no first order discontinuity in the quantities discussed above, which makes harder to localize the transition to the MI, since the orbitals are truly at half-filling only for $U \rightarrow \infty$. Nevertheless, around $U \sim 4$ we observe a change of concavity in the plotted quantities, which we can identify as the onset of the MI. \\
\noindent
In Fig.~\ref{fig:Figure2}  we plot $\Delta \hat{\Sigma}_{pp} ( i \epsilon, \mathbf{k}) =\text{Re} \hat{\Sigma}_{pp} ( i \epsilon, \mathbf{k}) - \hat{\Sigma}^{\text{HF}}_{pp}$ as a function of $\epsilon$ for different values of $U$, where $\hat{\Sigma}^{\text{HF}}_{pp}$ is the Hartee-Fock contribution, i.e. the $|\epsilon| \rightarrow \infty$ limit of the self-energy. We observe that $\Delta \hat{\Sigma}_{pp} ( i \epsilon, \mathbf{M}) = -\hat{\Delta} \Sigma_{ss} ( i \epsilon, \mathbf{M})$ looks like the difference of a very broad Lorentzian-like function and a narrower one that appears above $U = 3$ and grows with further increase of $U$, while it is the sum of two Lorentzians if we look at the patch $\mathbf{\Gamma}$. 
Therefore, we fit the frequency dependence with two Lorentzian functions
\beal
\Delta \hat{\Sigma}_{pp}(i\epsilon,\mathbf{k}) = - \fract{ \alpha_\bk\,\gamma^2_{ \mathbf{k} } }{ \;\epsilon^2 + \alpha^2_{ \mathbf{k} }\; } - \frac{ \epsilon_{ \mathbf{k} }\,\Delta^2_{ \mathbf{k} } } {\;\epsilon^2 + \epsilon_{\mathbf{k}}^2\;}
\label{fit}
\eal
where the first Lorentzian is smooth, broad and qualitatively the same in the two patches while the second one is narrower and opposite in sign in the two patches. We associate the latter with the emergence of a pole-like structure in the self-energy that becomes dominant in the MI phase. Looking at the fit parameters of the latter we can identify two trends: first, $\Delta_{ \mathbf{M} }^2$ increases making the contribution due to this Lorentzian dominant in the MI; second, $\epsilon_{ \mathbf{M} }$ changes sign between $U =5$ and $U =6$. To rationalize such a behaviour we consider a toy model where the effect of the broader Lorentzian is captured in a renormalization of the effective on-site mass $M$ and for the frequency dependent part of the self-energy we retain only the singular Lorentzian. Then the renormalized energies are given by
\beal
\hat{R}( \mathbf{k} ) = \hat{H}_{0}( \mathbf{k} ) +  \hat{\Sigma}_1(0, \mathbf{k} ) = \hat{H}_{0}( \mathbf{k} ) + \frac{ \Delta^2 }{ \hat{H}_1( \mathbf{k} ) }\;,
\label{toy-R}
\eal
where $\hat{H}_0(\bk)$ has an effective mass $M_{ \text{eff} } < M $, and $\hat{H}_1( \mathbf{k} )$ resembles $\hat{H}_0( \mathbf{k} )$ in 
\eqn{H_0}  
though with renormalized parameters. The change of sign of $\epsilon_{ \mathbf{M} }$ corresponds to a change of the mass $M_1$ in $\hat{H}_1(\bk)$ from $-2 t_1 < M_1 < 0$ to $M_1 < - 2 t_1$, i.e. a topological phase transition of the dispersive poles of the self-energy through a Dirac-like Luttinger surface, actually a point at $\mathbf{M}$. Finally, we recall that we do not have direct access to the off-diagonal elements of $\Sigma(i\epsilon, \mathbf{k} )$ which will be non-zero away from the high-symmetry points. It has been shown \cite{Adriano-CDMFT-BHZ-PRB2021} that such terms do not play a crucial role in the QSHI-MI transition. Nonetheless, in the following discussion we add to 
$\hat{H}_1(\bk)$ in \eqn{toy-R} a small $\lambda_1 = 0.2$ to correctly reproduce the gapping of the Luttinger surfaces away from the high-symmetry points. \\
\begin{figure}[t!] 
\includegraphics[width=0.5\textwidth]{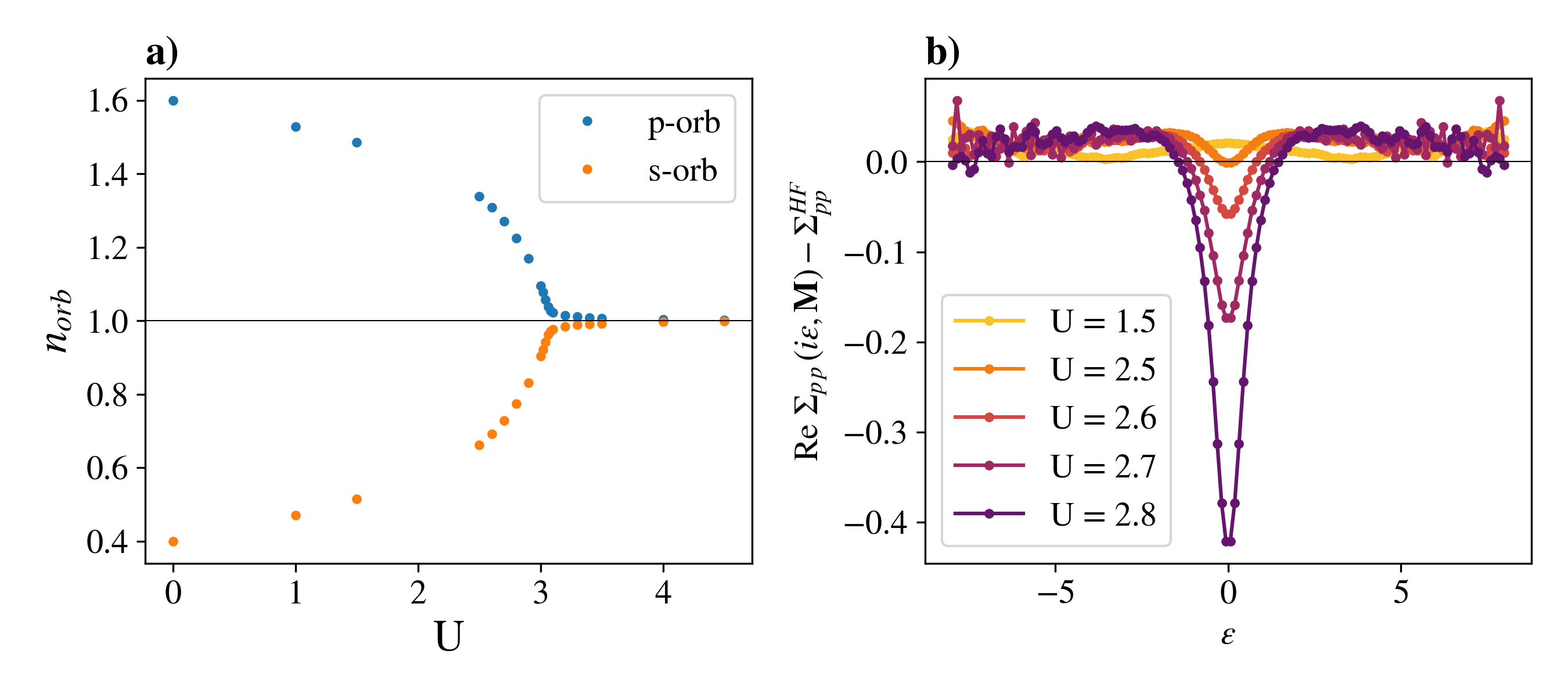}
\caption{\label{fig:Figure4}Panel a): Renormalization of the occupations in the two orbitals across the QSHI-MI transition, at fixed $M = 1.0$ and $J<0$. Panel b): $\Delta \hat{\Sigma}_{pp} ( i \epsilon, \mathbf{M}) = -\hat{\Delta} \Sigma_{ss} ( i \epsilon, \mathbf{M})$ for some values of $U$ before the onset of Mottness ($U\sim 3.1$). The phenomenology is the same already discussed in the $J>0$ case, as we can conclude comparing with Fig.~\ref{fig:Figure1} and Fig.~\ref{fig:Figure2}. }
\end{figure}
\noindent
In Fig.~\ref{fig:Figure3}, we plot the bands of the QP Hamiltonian $ \hat{H}_*(\bk)=\hat{H}_{*}(0, \mathbf{k} )$, see \eqn{H*}, both in periodic boundary conditions (PBC) and in open boundary conditions (OBC) along one direction \cite{note2}. For small values of the interaction (first panel) there is only a renormalization of the on-site mass difference, thus the system is adiabatically connected to the non-interacting QSHI and, in OBC, it has in-gap genuine chiral edge states, i.e., gapless surface Green's function poles. Increasing the interaction strength (second panel), dispersive poles appear in the self-energy ($\Delta \neq 0$), i.e., Green's function zeros, but, at first, without passing through a bulk gapless point, thus leaving unchanged the topological invariant. We observe that in OBC there are actually three pairs of chiral edge states: two associated with Green's function poles and one with zeros. This phase, in which there is a non-trivial interplay between the topological band of poles and zeros, has been recently observed \cite{Andrea&Ivan} in a model topological Kondo insulator and dubbed \textit{topological pseudogap insulator}.
Upon further increasing $U$ (third panel), the system undergoes a topological phase transition: the gap between the bands of poles closes at the high-symmetry points $\mathbf{X}$ and $\mathbf{Y}$. Beyond this semi-metal point, the model turns into a MI that has topological in-gap bands of zeros with winding numbers $W(G_\sigma)$ opposite to the QSHI, and not necessarily corresponding to finite Chern numbers $C_\sigma$ \cite{Andrea-Zeri-PRB2023}. In OBC we find only a pair of chiral edge zeros. Finally, deep inside the MI also the zeros become topologically trivial and the system can smoothly evolve towards the atomic limit (fourth panel). \\
\noindent
Finally, we want to emphasize that the continuous nature of the QSHI-MI transition remains robust even lowering the temperature, indicating that this is not merely a finite temperature effect but more likely the consequence of the appearance of Green's function bands of zeros already in the QSHI, which smoothly evolve in the MI upon increasing $U$.
\subsection{ The $J < 0$ case}
We now briefly discuss the symmetric QSHI-MI phase transition in the $J < 0$ case. We choose $ J = - \frac{U}{2} $ so that $V = 0$, which allows us to make a direct comparison with the bilayer model investigated in \cite{Slage-PRB2015} where the two layers interact solely via an antiferromagnetic exchange. As previously discussed, we also add the interaction term \eqn{interaction-dipole-1} to favour the on-site spin-singlet configuration, thus a low-spin MI (ls-MI). In our computation, we set $\delta J = U / 16$.  \\
In Fig.~\ref{fig:Figure4} we show the occupation numbers $n_{orb}$ of the two orbitals with increasing $U$ at fixed $M=1.0$, as well as $\Delta \hat{\Sigma}_{pp} ( i \epsilon, \mathbf{k}) =\text{Re}\, \hat{\Sigma}_{pp} ( i \epsilon, \mathbf{k}) - \hat{\Sigma}^{\text{HF}}_{pp}$ as a function of $\epsilon$ for different values of $U$. We find that the phenomenology of the QSHI-MI phase transition in this case is the same as for positive $J$, thus occurs through a closing of the single-particle gap. This is different from what was reported in \cite{Slage-PRB2015}, where a similar but different model was examined. Additionally, in their study, they set $|J|/U = 2$, meaning the transition is controlled by the $J$ term, whereas in our study the transition is driven by $U$.
\begin{figure}[t!] 
\includegraphics[width=0.5\textwidth]{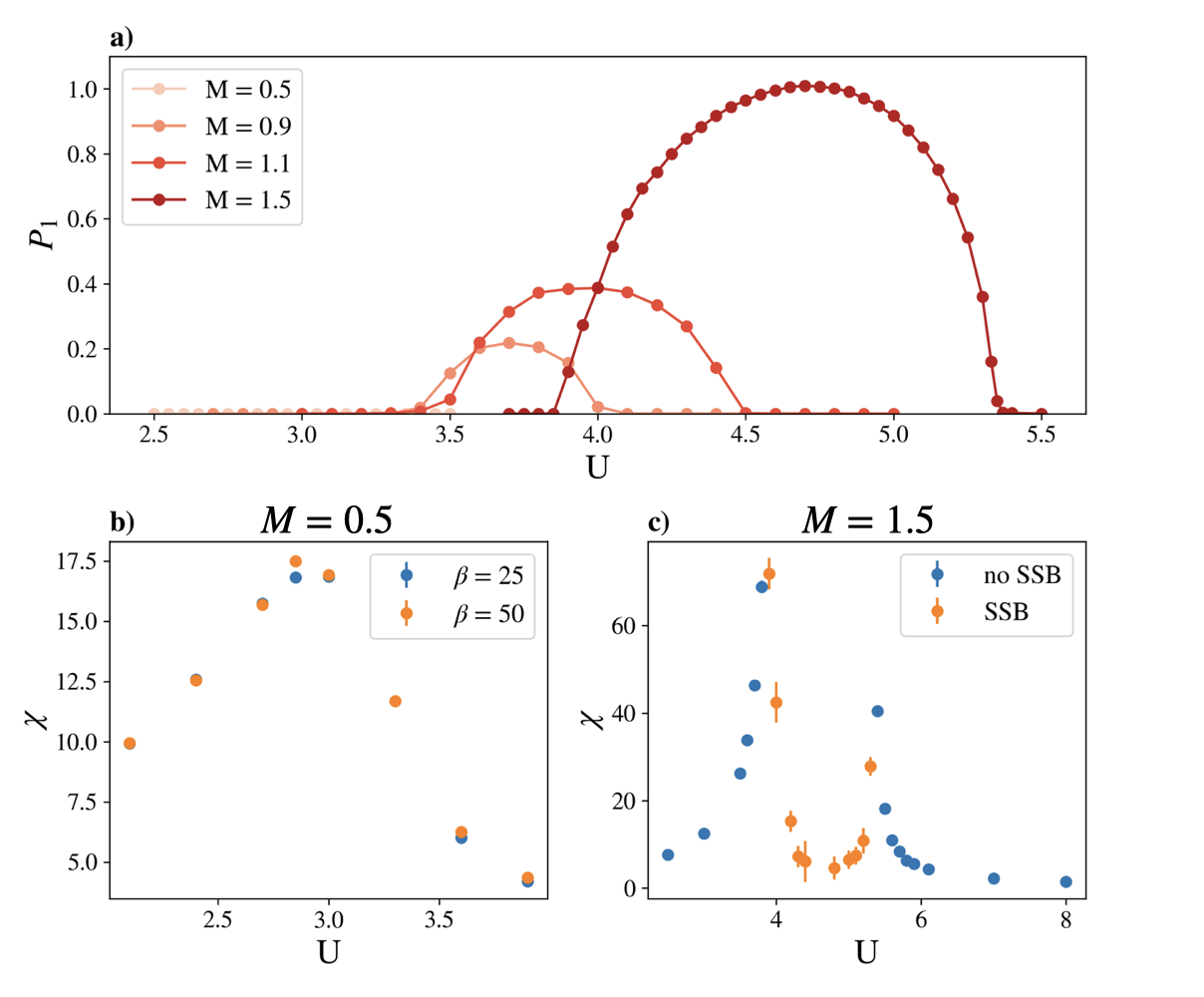}
\caption{\label{fig:Figure5}Panel a): Excitonic order parameter $P_1$ as a function of the interaction strength $U$ for 4 different values of $M$. As we decrease $M$ the bell-like dependence is strongly renormalized both in height and width, until for $M=0.5$ no condensation appears. The results we show are stable until the lowest temperature we checked, i.e. $\beta = 150$. Panel b): We support the claim that for small values of $M$ (e.g. $M=0.5$) no condensation appears with the computation of the susceptibility $\chi$ to an external field coupled to the excitonic order parameter. There is no substantial difference between $\beta =25$ and $\beta=50$, further suggesting this is not a temperature consequence. Panel c): For $M=1.5$ the susceptibility diverges at the phase boundaries of the EI.}
\end{figure}
\section{The excitonic instability}
\subsection{ The $J > 0$ case}
We now investigate what could happen if we allow for spontaneous symmetry breaking (SSB) in the system. In particular, we focus on the possible emergence of an exciton condensate characterized by the order parameter \eqn{exciton J>0}, e.g., at $\phi = 0$, 
\beal
P_{1} = \langle \Psi^\dagger_{\bR} \, \sigma_1\otimes\tau_1\, \Psi_{\bR} \, \rangle\,.
\label{P1}
\eal
We call excitonic insulator (EI) the phase where $P_1 \neq 0$, which breaks the symmetries that protect the non-trivial topology of the QSHI. In Fig.~\ref{fig:Figure5} we plot $P_1$ as a function of the interaction strength for different values of $M$. Both the QSHI-EI and EI-MI transitions seem to be  second order within 2-patches DCA. We recall that in single site DMFT at $T=0$ only the latter is so, while the QSHI-EI one is first order \cite{Adriano-PRB2023}. For smaller values of $M$, the range of stability of the EI shrinks. This could already be observed in DMFT, though the EI survives down to $M \rightarrow 0$. On the contrary, the non-local correlations captured in DCA seem to prevent the stabilisation of such a phase for $M \lesssim 0.6$, at least down to the lowest accessible temperatures. This scenario is further supported by the direct computation at $M= 0.5$ of the static and uniform susceptibility $\chi$ to a field coupled to the excitonic order parameter, see Fig.~\ref{fig:Figure5}. In contrast to what we observe at larger $M$, e.g. $M = 1.5$, there is no value of $U$ such that the susceptibility diverges. However, we cannot exclude that, at lower values of temperatures than those we can access, the EI does appear. 
\begin{figure}[t!] 
\includegraphics[width=0.4\textwidth]{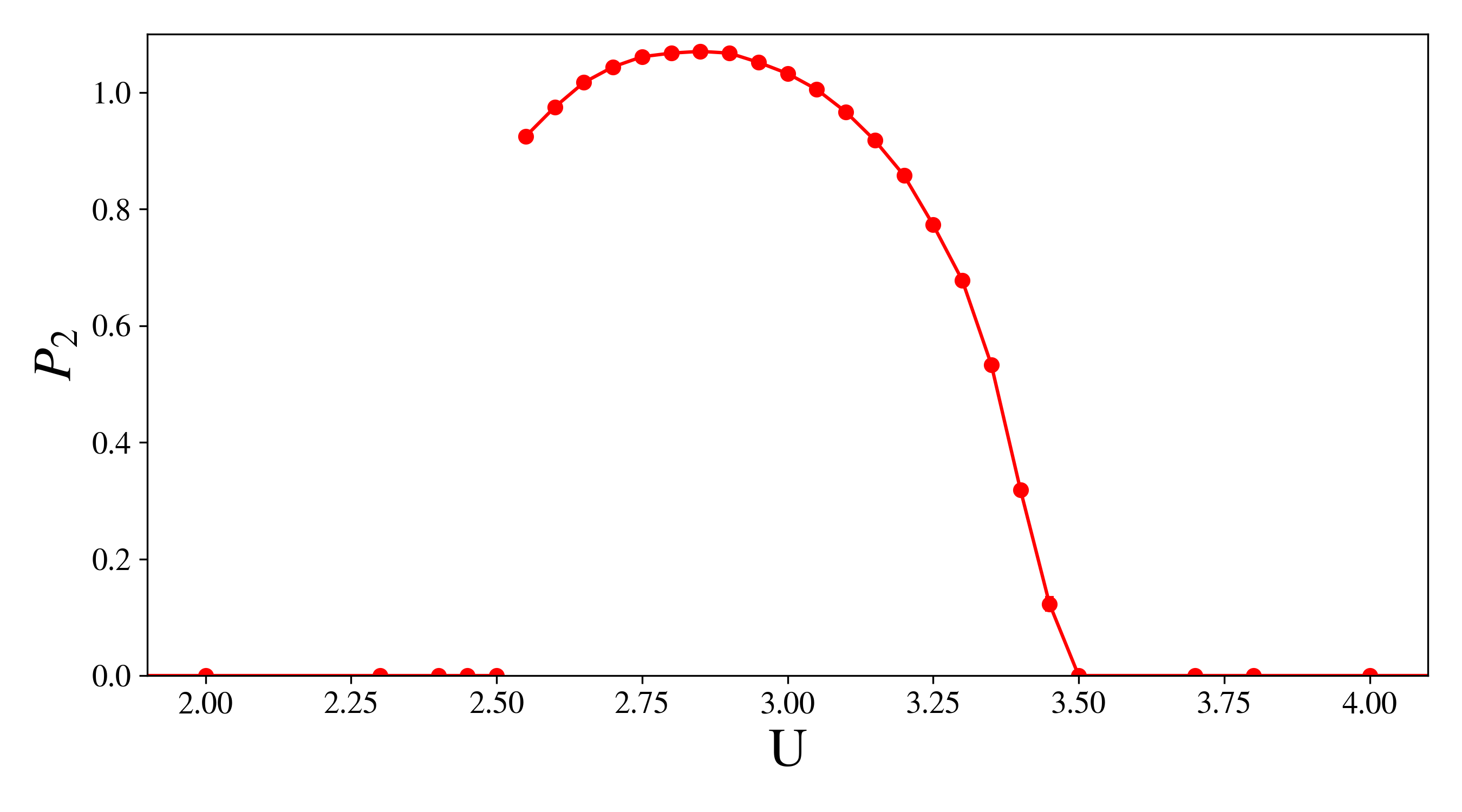}
\caption{\label{fig:Figure6}Excitonic order parameter $P_2$ in the $J<0$ case for $M=1.0$. Differently from the $J>0$ case the QSHI-to-EI transition is first order, while the EI-to-MI is still second order.}
\end{figure}
\subsection{ The $J < 0$ case}
In this case the EI phase is characterized by an excitonic order parameter \eqn{exciton J<0}, 
\beal
P_{2} = \langle \Psi^\dagger_{\bR} \, \sigma_0\otimes\tau_1\, \Psi_{\bR} \, \rangle\,,
\label{P2}
\eal
breaking $\mathcal{I}$, but not $\mathcal{T}$, see Fig.~\ref{fig:Figure6}. However, unlike 
for $J>0$, the QSHI-EI transition looks first order, while, the EI-MI transition is again second order. We notice that also with negative $J$, we cannot stabilize an EI at small values of $M$, at least down to the lowest numerically accessible temperature.  
\begin{figure}[t!] 
\includegraphics[width=0.5\textwidth]{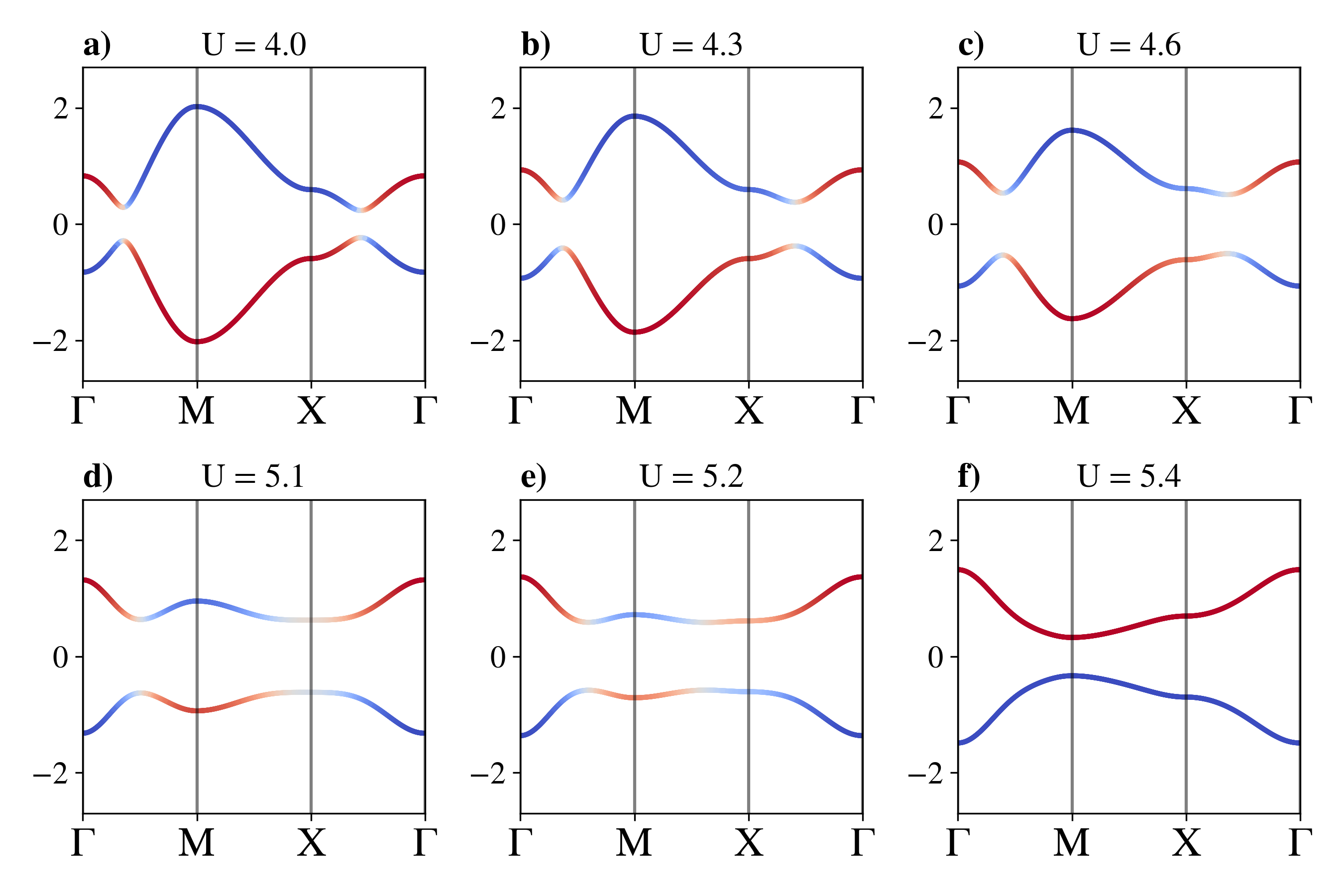}
\caption{\label{fig:Figure7}Smooth evolution of the QP bands across the QSHI-EI-MI transitions. The colours of the lines represent the orbital character of the eigenvectors, blue(red) for the s-(p-)character. In panel a) the system is a QSHI. In panels b), c), d) and e) the system is an EI with $P_1 \neq 0$. The EI breaks inversion symmetry, indeed the orbital character at the high-symmetry points is no more well defined. In panel f) the QP bands closely resemble the dispersive Green's function zeros of a MI.}
\end{figure}
\subsection{ Excitonic instability of the Mott insulator and the quasiparticle scenario }
Here, we try to shed some light on another intriguing issue related to the MI-EI transition, 
focusing for simplicity on the $J>0$ case. Indeed, while the QSHI-EI transition can be conceptually grasped through the softening of an exciton bound state between valence and conduction bands as the single-particle gap shrinks upon increasing $U$, this picture is not justifiable at the MI-EI transition. In the MI, the single-particle gap typically remains large, seemingly precluding the formation of excitonic bound states capable of softening upon approaching the transition. This quite natural observation prompted us to explore whether in the MI the excitons can be regarded as bound states of the valence and conduction bands of Green's function zeros. The idea comes spontaneously once we notice that a Luttinger surface may host Landau's quasiparticles, similarly to a Fermi surface \cite{Luttinger_surface_QP}. These quasiparticles cannot transport charge if the system is a true Mott insulator, but they are not neutral in the strict sense. In other words, one should imagine \cite{Note1}
that the quasiparticle motion in the MI is cancelled by a charge backflow so that there is no net longitudinal current flowing in an electric field. On the contrary, such dressed quasiparticle could still transport entropy, spin, and even experience a Lorentz force in a magnetic field \cite{mio-Mott,Andrea&Ivan}. In our case, the quasiparticle carries an additional quantum number, the parity, 
which not only makes it possible the opening of a hybridisation gap, but also the existence of 
bound states between opposite-parity quasiparticle and quasihole excitations. Since these anomalous 
quasiparticles are closely related to in-gap bands of Green's functions zeros \cite{Andrea-Zeri-PRB2023,Giorgio-NatComm2023,wagner2023edgezerosboundaryspinons}, we can indeed look upon their bound states as those between valence and conduction bands of Green's functions zeros. Interestingly enough, it has been shown in a similar model that a MI with a Luttinger surface has an odd-parity exciton susceptibility that is power-law in temperature, suggestive of the existence of gapless opposite-parity quasiparticles \cite{Andrea&Ivan}. \\
\noindent
Moving back to our results, in Fig.~\ref{fig:Figure7} we show the QP bands across the two transitions for $M = 1.5$ and $J>0$. In the following we will refer to $U_{c1}$ ($U_{c2}$) the value correspondent to the QSHI-EI (EI-MI) transition. For $U< U_{c1}\simeq 3.9$, the system is a QSHI. For $U \in [ U_{c1}, U_{c2}]$, with 
$U_{c2}\simeq 5.3$, we instead find the EI with 
$P_1 \neq 0 $. As we mentioned, $P_1 \neq 0$ breaks inversion and, indeed, 
the orbital character at the high-symmetry points is no more well defined, 
see Fig.~\ref{fig:Figure7}. Finally, for $U>U_{c2}$ the QP bands closely resemble the dispersive Green's function zeros of a MI. We observe that the QP bands evolve smoothly across the two very different transitions, which is a advantageous feature of the quasiparticle description outlined in the equations 
\eqn{G*}-\eqn{H*}\,.
\begin{figure}[t!] 
\includegraphics[width=0.45\textwidth]{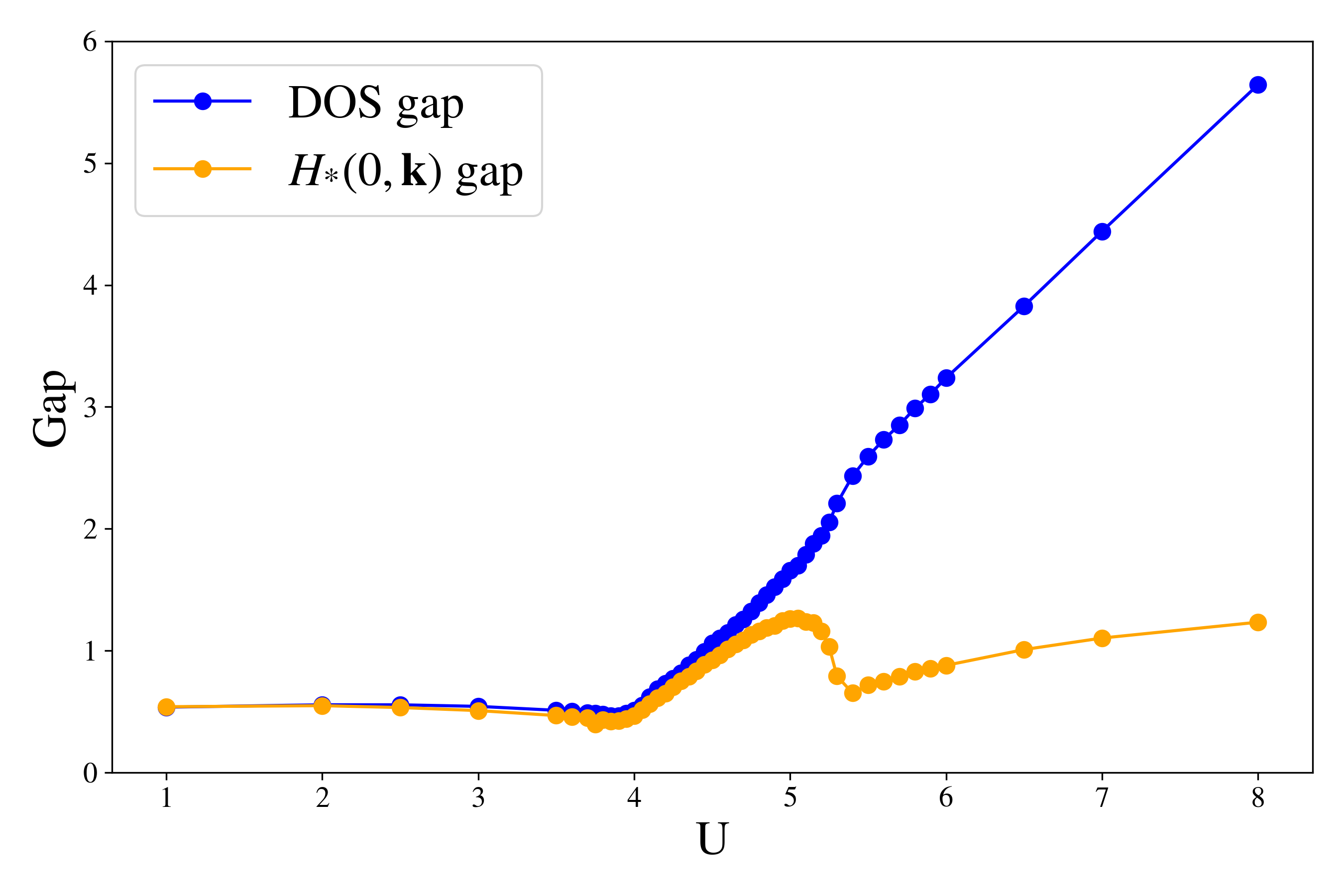}
\caption{\label{fig:Figure8}Panel a): Comparison between the QP gap of $H_{*}(0, \mathbf{k})$ versus the single-particle one extracted from the DOS in the QSHI-EI-MI transitions. The point at which the two start to deviate signal that the bands of zeros come into play. We notice that in the MI phase the QP gap is much smaller than the DOS one and, as expected the condensation of excitons increases the QP gap due to the additional level repulsion between valence and conduction bands.}
\end{figure}

\noindent
In Fig.~\ref{fig:Figure8} we show the QP gap, i.e., the gap of the QP Hamiltonian $H_*(\bk)$, versus the single-particle one extracted from the density-of-states (DOS).
We emphasise that, in conventional insulators, the condensation of excitons is expected  
to raise the single-particle gap, since it generates an additional level repulsion between valence and conduction bands. This is indeed the case at the QSHI-EI transition, but in no way it occurs 
at the MI-EI transition, where the single-particle gap actually decreases, see Fig.~\ref{fig:Figure8}. 
On the contrary, the QP gap does precisely what is expected: it grows entering the EI phase from both 
the QSHI and MI sides. We believe that this remarkable behaviour is a strong indication that the 
exciton that becomes soft in the MI is formed between valence and conduction bands of zeros rather than 
from the lower and upper Hubbard bands. This is further supported by the evidence that the single-particle 
gap in the MI is sizeable, which makes not conceivable that lower and upper Hubbard bands 
might be involved in the exciton softening. We end noticing that single-particle and QP gaps are practically the same on increasing $U$ up to around the value at which $P_1$ reaches a maximum, see Fig.~\ref{fig:Figure5}. Above that, while the single-particle gap keeps growing with $U$, the QP one first decreases up to the Mott transition and then it raises again, though remaining much smaller than the single-particle gap. The point at which the two gaps start to deviate from each other is also that when the bands of zeros come into play, which thus occurs already in the EI. Indeed, looking at the low-frequency behavior of the self-energy in the EI we can still identify a structure made up of two Lorentzians, as the one we discussed in Fig.~\ref{fig:Figure2}.

\section{Conclusions}
We have extensively analysed the interacting BHZ model by means of dynamical cluster approximation \cite{DCArev}, 
which allows us to study in detail the intriguing interplay between the bands of Green's function poles 
and those of Green's function zeros upon approaching the Mott transition \cite{Gurarie-zeros-PRB2011,Essing&Gurarie-PRB2011}. 
We have discovered that bands of Green's function zeros emerge already in the quantum spin-Hall insulator, 
leading to a rather remarkable topological insulator exhibiting two chiral branches of edge Green's function poles and one of zeros, in that way leaving invariant the spin Chern number of the weakly correlated state \cite{Andrea&Ivan} and thus without requiring a topological transition. \\
The Mott transition forcing symmetries occurs as usual through a semi-metal point where the single-particle gap closes at the high-symmetry $\mathbf{X}$ and $\mathbf{Y}$ points, though with a vanishing quasiparticle residue that entails a jump of the single-particle gap to the large separation 
between lower and upper Hubbard bands. However, inside the Mott insulator valence and conduction topological bands 
of Green's function zeros survive and induce, in open boundary conditions, edge zeros. At a critical 
value of the interaction strength the bands of zeros touch in a Dirac-like manner at the high-symmetry point $\mathbf{M}$, and, above such value, these bands become trivial and the Mott insulator is able to evolve smoothly into its atomic limit \cite{Giorgio-NatComm2023}. \\
\noindent
When we allow for symmetry breaking, a non-topological excitonic insulator is found to intrude between 
the quantum spin-Hall and Mott insulators \cite{Blason2020PRB,Adriano-PRB2023}. We do not find 
the expected magnetic instability in the Mott insulator, due to a well known flaw of  
the two-patch dynamical cluster approximation calculation \cite{Maier-PRL2005} that we can afford. 
However, to ensure that our results are not an artifact of the approximation, we have also studied a model with an inverted exchange that stabilises a Van Vleck paramagnetic Mott insulator. Indeed, the 
two models show a similar behavior. In particular, we find evidence that the exciton in the 
Mott insulator, which becomes soft at the transition into the excitonic insulator, might actually be a bound state between valence and conduction bands of Green's function zeros rather than between lower 
and upper Hubbard bands, which are very distant in energy.

\section*{Acknowledgments}
We are very grateful to Adriano Amaricci and Carlos Mejuto Zaera for helpful discussions and comments.


\begin{thebibliography}{31}
\bibitem [{not()}]{note2}%
  \BibitemOpen
  \href@noop {} {}\bibinfo {note} {In the OBC geometry we used the fit
  parameters of the results in PBC to construct the hopping processes in
  $H_{0}(\mathbf{k})$ and $H_{1}(\mathbf{k})$. Starting from that we construct
  the matrices of $H_{0}(\mathbf{k})$ and $H_{1}(\mathbf{k})$ with PBC along x
  and OBC with $N_y = 200$ layers along y.}\BibitemShut {Stop}%
\end{thebibliography}
\end{document}